\documentclass[10pt,aps,prx,twocolumn,amsmath,amssymb,showpacs,superscriptaddress]{revtex4-2}

\usepackage{amsmath,amssymb,amsfonts}
\usepackage{graphicx}
\usepackage{bm}
\usepackage[colorlinks=true, linkcolor=blue, citecolor=green, urlcolor=magenta]{hyperref}
\usepackage{tikz}

\usepackage{xcolor}

\begin{document}

\sloppy
\title{Cycle holonomy captures higher-order compatibility constraints in remote synchronization}

\author{Llu\'is Torres-Hugas}
\affiliation{Departament d'Enginyeria Inform\`{a}tica i Matem\`{a}tiques, Universitat Rovira i Virgili, 43007 Tarragona, Spain}

\author{Jordi Duch}
\affiliation{Departament d'Enginyeria Inform\`{a}tica i Matem\`{a}tiques, Universitat Rovira i Virgili, 43007 Tarragona, Spain}
\affiliation{Center for Computational Science and Applied Mathematics (ComSCIAM), Universitat Rovira i Virgili, 43007 Tarragona, Spain}

\author{Sergio G{\'o}mez}
\affiliation{Departament d'Enginyeria Inform\`{a}tica i Matem\`{a}tiques, Universitat Rovira i Virgili, 43007 Tarragona, Spain}
\affiliation{Center for Computational Science and Applied Mathematics (ComSCIAM), Universitat Rovira i Virgili, 43007 Tarragona, Spain}

\author{Alex Arenas}
\email{alexandre.arenas@urv.cat}
\affiliation{Departament d'Enginyeria Inform\`{a}tica i Matem\`{a}tiques, Universitat Rovira i Virgili, 43007 Tarragona, Spain}
\affiliation{Center for Computational Science and Applied Mathematics (ComSCIAM), Universitat Rovira i Virgili, 43007 Tarragona, Spain}

\begin{abstract}
Higher-order interactions have typically been modeled using hypergraphs or simplicial complexes, where interactions explicitly involve more than two nodes. Here we demonstrate that effective higher-order dynamical constraints emerge naturally on ordinary graphs, provided the interaction carries nontrivial topological structure. We study a gauge-coupled phase model with edge phase lags whose accumulation around closed loops produces gauge-invariant mismatches. We show that the associated twisted Laplacian admits a zero mode if and only if all cycle holonomies vanish. Consequently, global compatibility is obstructed not by local pairwise mismatches, but by intrinsic topological frustration on cycles. We then connect this framework to the symmetric Sakaguchi--Kuramoto model, whose local coupling law differs from the gauge-coupled model but whose node phases impose cycle closure on physical phase differences. For cactus graphs, path mismatches induced by the symmetric lag can be represented through associated cycle holonomies, providing a static spectral encoding of their global residual incompatibility. Our results establish a spectral framework linking frustration to cycle-level constraints and identify cycle holonomy as a local-to-global diagnostic of path incompatibility in synchronization dynamics.
\end{abstract}

\maketitle

\section{Introduction}

Collective behavior in physical and biological systems is often shaped by constraints that cannot be understood by considering each edge independently.~\cite{strogatz2000kuramoto,arenas2008synchronization,dorfler2014synchronization}. To capture these complexities, higher-order interaction networks, such as hypergraphs and simplicial complexes, have become a central framework, successfully describing phenomena like multistability, clustered synchronization, and explosive transitions  \cite{battiston2020beyond,millan2020explosive,battiston2021collective,carletti2024diracsync,wang2023higherorderglobalsync}.

Here we focus on remote synchronization \cite{bergner2012remote,gambuzza2013analysis}, a counterintuitive phenomenon where nodes synchronize despite being separated by intermediate, incoherent nodes.  While often attributed to specific symmetry breaking or frustration mechanisms  \cite{nicosia2013remote,pecora2014cluster,sorrentino2016complete}, we propose that remote synchronization can be understood as a manifestation of effective higher-order constraints inherent in the network's topology. Crucially, we show that these constraints arise even in ordinary graphs, provided the interaction supports a nontrivial gauge structure.

We develop a spectral framework based on the twisted Laplacian. We assign a phase lag to each oriented edge and use it to compare node phases along the network. In this geometric picture, node states are compared after transport along edges. The resulting energy functional defines a twisted Laplacian whose spectrum encodes the global compatibility of the phase-transport problem~\cite{shubin1994dml,singer2012vdm,bandeira2013cheeger,gao2021geometry,hansen2019toward,hansen2021opinion}.

Although phase transport is defined locally, its consequences are global. The obstruction to synchronization is determined by the holonomy, that is, the accumulated phase shift along closed loops. If the holonomy is trivial on every cycle, the phase lags are gauge-trivial and can be eliminated through a change of variables at the nodes. If at least one cycle has nontrivial holonomy, the local phase relations are globally incompatible, producing intrinsic topological frustration that cannot be resolved edge by edge. This dependence on the collective structure of a cycle is the sense in which higher-order constraints emerge, even though the coupling remains pairwise.

We formalize this by constructing the twisted incidence operator $\delta_\alpha$ and its Laplacian $L_\alpha=\delta_\alpha^{*}\delta_\alpha$. We show that $L_\alpha$ possesses a zero mode if and only if all cycle holonomies vanish. This establishes a correspondence between synchronization compatibility and cycle closure. Furthermore, we show that the smallest eigenvalue of $L_\alpha$ provides a spectral measure of global residual incompatibility.

Within this framework, the gauge-coupled phase dynamics arises as the gradient descent of the twisted energy on the torus, in a co-rotating frame. The low-lying eigenmodes of the twisted Laplacian describe the phase configurations that are least affected by the incompatibility introduced by nontrivial cycle holonomy. We then use this spectral structure as a static reference for the symmetric Sakaguchi--Kuramoto model~\cite{sakaguchi1986active}, where remote synchronization emerges from the nonlinear phase dynamics. For cactus graphs, the path mismatches generated by the symmetric phase lag can be represented through associated cycle holonomies. This connects our construction to the broader role of cycle constraints in phase-locked oscillator networks~\cite{manik2017cycle}, and allows the twisted Laplacian spectrum to provide a global spectral description of the resulting compatibility structure.

For a constant phase lag $\alpha$ on cycle-based motifs, the relevant cycle-level quantity is the effective mismatch $h(\alpha)$ generated by phase accumulation along competing paths. For an $\ell$-cycle substructure, this mismatch is given by $h(\alpha)=(\ell-2)\alpha \pmod{2\pi}$, as we will show later. For an isolated cycle, the two lowest twisted spectral branches become degenerate when the mismatch reaches the maximally frustrated value $h(\alpha_\pi)=\pi$, which gives $(\ell-2)\alpha_\pi=\pi$. This condition identifies the point at which the competing paths are maximally incompatible and provides a spectral marker against which reorganizations of the phase-locked dynamics can be compared. For a $5$-cycle substructure, it yields $\alpha_\pi=\pi/3$, close to the characteristic value previously associated with the remote synchronization transition in Ref.~\cite{nicosia2013remote}, but here identified as the point of maximal path incompatibility rather than a stability boundary.

In short, we show how higher-order dynamical constraints can arise on ordinary graphs once the coupling carries a gauge structure. In this setting, cycle holonomy is the relevant gauge-invariant obstruction, and the low-lying spectrum of the twisted Laplacian links topology to phase compatibility. Remote synchronization patterns can then be related to an effective cycle mismatch. For an isolated cycle, when this mismatch reaches $\pi$, the lowest twisted eigenmodes reorganize, marking the point of maximal path incompatibility and providing a spectral reference for the reorganization of the phase-locked dynamics.

\section{Results}

\subsection{Local phase lags and gauge-invariant cycle mismatch}

We begin by introducing a coupling structure that extends standard diffusive coupling on graphs while preserving its pairwise character. Let $G=(V,E)$ be a connected graph with $N$~nodes, and fix an arbitrary orientation on each edge. To every oriented edge $i\to j$ we assign a phase transport factor
\begin{equation}
U_{ij}
=
e^{i\alpha_{ij}},
\qquad
\alpha_{ji}
=
-\alpha_{ij},
\label{cond}
\end{equation}
so that reversing the orientation inverts the transport. The collection $\alpha=\{\alpha_{ij}\}$ specifies how phases are compared across edges.

In the standard Kuramoto setting, phase differences are measured directly as $\phi_j-\phi_i$. Here, however, before comparing node phases, the state at node $i$, written as $z_i=e^{i\phi_i}\in\mathbb{C}$, is parallel-transported along the edge and compared with node $j$, that is, $z_j$ is compared with $U_{ij}z_i$. In this sense, $\alpha_{ij}$ acts as a discrete parallel transport encoding an edge-dependent phase shift.

To analyze this coupling within our spectral framework, we assign $\alpha_{ij}$ along a fixed reference orientation for each edge and impose $\alpha_{ji}=-\alpha_{ij}$. The reference orientation is only a bookkeeping convention: reversing it also reverses the sign of the corresponding phase lag and leaves the underlying connection unchanged. The signed phase assigned to each edge, however, is part of the connection. In the second part of the paper, these phases are chosen so that their cycle holonomies reproduce the path mismatches induced by the symmetric Sakaguchi--Kuramoto lag.

Although the transport is defined locally on edges, its global compatibility is controlled by how these phase shifts accumulate along closed loops. For any oriented cycle $C$, the accumulated phase mismatch
\begin{equation}
\mathrm{Hol}_\alpha(C)
=
\sum_{(i,j)\in C} \alpha_{ij}
\quad
\pmod{2\pi}
\end{equation}
is invariant under gauge transformations. This is the holonomy of the connection around $C$ and provides a cycle-level measure of frustration. Reversing the orientation of the cycle changes its sign but not whether it vanishes. If this accumulated phase vanishes for every cycle, the edge phase lags can be removed by redefining node phases. If not, the mismatch cannot be eliminated globally, and the system is intrinsically frustrated.

This structure is naturally expressed in terms of gauge transformations. A node-wise phase redefinition
\begin{equation}
z_i
\mapsto
e^{i\theta_i} z_i
\end{equation}
induces a transformation
\begin{equation}
\alpha_{ij}
\mapsto
\alpha_{ij} + \theta_j - \theta_i.
\end{equation}
Two collections of edge phases related in this way are gauge-equivalent. What remains invariant under such transformations is completely determined by the holonomies on a cycle basis. On trees, where no cycles exist, every phase lag configuration can be removed by a gauge transformation. In contrast, on graphs with nontrivial cycle structure, nonvanishing holonomy represents a genuine global obstruction.

\subsection{From compatibility energy to gauge-coupled phase dynamics}
\label{sec:sec2.2}

To obtain a linear algebraic formulation, we first relax the phase-only description and represent each oscillator by a complex state $z_i\in\mathbb{C}$. This is convenient because phase shifts act multiplicatively as rotations in the complex plane. Our goal is to quantify how well neighboring oscillators satisfy the prescribed coupling constraint once the phase lag is taken into account. Along an oriented edge $i\to j$, we compare $z_j$ with the transported state of $z_i$, which defines the twisted incidence operator
\begin{equation}
(\delta_\alpha z)_{ij}
=
z_j-e^{i\alpha_{ij}}z_i.
\end{equation}
When $\alpha_{ij}=0$, this reduces to the standard graph incidence operator $z_j-z_i$, so the present construction directly generalizes ordinary diffusive coupling~\cite{hansen2019toward}.

The quantity $(\delta_\alpha z)_{ij}$ measures the local incompatibility between neighboring states once the prescribed phase transport has been taken into account. A natural way to quantify global incompatibility over the entire network is to sum the squared magnitudes of these edge mismatches. The total cost of this incompatibility is the quadratic energy
\begin{equation}
\mathcal{E}(z)
=
\frac12 \sum_{(i,j)\in E}
\left|
z_j - e^{i\alpha_{ij}} z_i
\right|^2
=
\frac12 \|\delta_\alpha z\|^2.
\end{equation}
The factor $1/2$ is conventional and has no dynamical consequence.

This energy penalizes deviations from compatibility under the given connection. For a given state $z$, it vanishes precisely when every edge constraint
\begin{equation}
z_j=e^{i\alpha_{ij}}z_i
\end{equation}
is satisfied. Thus, the connection is globally compatible if and only if there exists a nonzero state $z$ for which $\mathcal{E}(z)=0$, so that all transported states match exactly throughout the network.

The quadratic structure of $\mathcal{E}(z)$ naturally defines the linear operator
\begin{equation}
L_\alpha
=
\delta_\alpha^{*}\delta_\alpha,
\end{equation}
which generalizes the standard graph Laplacian, where $\delta_\alpha^{*}$ denotes the Hermitian adjoint of the twisted incidence operator. By construction,
\begin{equation}
\mathcal{E}(z)
=
\frac{1}{2}
\langle z,L_\alpha z\rangle.
\end{equation}
The corresponding Euler--Lagrange equation is
\begin{equation}
L_\alpha z
=
0.
\label{ELeq}
\end{equation}
Since the trivial solution $z=0$ always exists, global compatibility is characterized by the existence of a nonzero solution of Eq.~\eqref{ELeq}. Thus, $\ker L_\alpha\neq\{0\}$ if and only if the edge compatibility conditions can be satisfied globally in the relaxed linear setting.

When all cycle holonomies vanish, Eq.~\eqref{ELeq} admits nontrivial solutions. If some cycle has nonzero holonomy, then $\ker L_\alpha=\{0\}$ and the smallest eigenvalue of $L_\alpha$ is strictly positive.

Finally, we impose the physical constraint that the oscillators have unit amplitude, $|z_i|=1$, so that $z_i=e^{i\phi_i}$. Substituting this into the energy yields
\begin{equation}
\mathcal{E}(\phi)
=
\sum_{(i,j)\in E}
\left[
1-\cos(\phi_j-\phi_i-\alpha_{ij})
\right].
\end{equation}
Its gradient descent gives a Sakaguchi--Kuramoto-type flow in a co-rotating frame, or equivalently for identical intrinsic frequencies,
\begin{equation}
\dot{\phi}_i
=
\kappa\sum_{j\sim i}
\sin(\phi_j-\phi_i-\alpha_{ij}),
\end{equation}
where $\kappa$ is the coupling strength.

Although the nonlinear dynamics evolves on a $N$-dimensional torus $ \mathbb T^{N} = \mathbb S^1 \times  \cdots \times \mathbb S^1$, the spectral structure of $L_\alpha$ remains informative. The eigenvectors associated with small eigenvalues describe the modes that best accommodate the incompatibility induced by nonzero cycle holonomy.

A particularly important case is a uniform phase lag,
\begin{equation}
\alpha_{ij}=\alpha
\end{equation}
for every edge $i\to j$ in the chosen reference orientation. To satisfy $\alpha_{ji}=-\alpha_{ij}$, traversing the same edge in the reverse direction contributes $-\alpha$. This convention defines a uniform antisymmetric connection, whose cycle holonomies are obtained by summing the signed edge contributions along each oriented cycle. This uniform antisymmetric connection defines a reciprocal gradient system, whereas the symmetric Sakaguchi--Kuramoto model uses the same phase lag in both interaction directions.

\subsection{Cycle compatibility and remote synchronization}

We now connect the gauge-coupled construction to the symmetric Sakaguchi--Kuramoto model. The two models have different local coupling laws. In the symmetric model, the same phase lag is used in both interaction directions, whereas in the gauge-coupled model reversing an edge inverts the transport factor. The correspondence therefore occurs at the level of path compatibility. The symmetric phase lag generates cycle-dependent path mismatches, and an associated antisymmetric connection represents these mismatches as cycle holonomies.

\subsubsection{Cycle mismatch in the symmetric Sakaguchi--Kuramoto model}

Consider the symmetric Sakaguchi--Kuramoto dynamics
\begin{equation}
\dot{\phi}_i
=
\omega
+
\kappa\sum_{j=1}^{N}A_{ij}
\sin(\phi_j-\phi_i-\alpha),
\label{eq:SK}
\end{equation}
where the same phase lag $\alpha$ appears in both interaction directions.

Let $C$ be a simple cycle of length $\ell$, and choose two adjacent nodes $u$ and $v$. These nodes are connected by the direct edge $e=(u,v)$ and by the complementary path $P_{uv}$ containing the remaining $\ell-1$ edges of the cycle. We traverse both paths from $u$ to $v$. Along the direct edge, the coupling argument is
\begin{equation}
\phi_v-\phi_u-\alpha.
\end{equation}
Summing the coupling arguments along the complementary path gives
\begin{equation}
\sum_{(i,j)\in P_{uv}}
\left(\phi_j-\phi_i-\alpha\right)
=
\phi_v-\phi_u-(\ell-1)\alpha,
\end{equation}
because the phases of the intermediate nodes cancel.

The mismatch between the coupling argument along the direct edge and the accumulated coupling argument along the complementary path is
\begin{align}
h_\alpha(C)
&=
(\phi_v-\phi_u-\alpha)
-
\left[
\phi_v-\phi_u-(\ell-1)\alpha
\right]
\nonumber\\
&=
(\ell-2)\alpha
\pmod{2\pi}.
\end{align}
This quantity is independent of the phase configuration and defines a phase-independent comparison between the coupling argument on the direct edge and the sum of the coupling arguments along the complementary path.

The selected edge is not physically distinguished. Every edge of a simple cycle has a complementary path containing $\ell-1$ edges, so the same mismatch
\begin{equation}
h_\alpha(C)
=
(\ell-2)\alpha
\pmod{2\pi}
\end{equation}
is obtained regardless of which edge is selected. Since a simple cycle has cycle rank one, this scalar determines the gauge class of an auxiliary connection on that cycle once its orientation is fixed.

\subsubsection{Path-holonomy correspondence on cactus graphs}

The cycle mismatch derived above can be represented as the holonomy of an associated antisymmetric connection. This construction applies directly to cactus graphs, where every edge belongs to at most one simple cycle. Equivalently, distinct cycles may share vertices but not edges. The mismatch associated with each cycle can therefore be encoded independently, without assigning conflicting connection phases to a shared edge.

Let
\begin{equation}
\mathcal{C}
=
\{C_1,\ldots,C_R\}
\end{equation}
denote the set of simple cycles of a connected cactus graph, and let $\ell_r=|C_r|$ be the length of cycle $C_r$. The symmetric Sakaguchi--Kuramoto interaction generates the path mismatch
\begin{equation}
h_\alpha(C_r)
=
(\ell_r-2)\alpha
\pmod{2\pi}
\end{equation}
between a direct edge and its complementary path around $C_r$.

To obtain an holonomy $\mathrm{Hol}_\alpha(C)$ of an associated antisymmetric connection that represents the discussed path mismatch $h_\alpha(C_r)$, we proceed as follows.  We choose an edge $e_r=(u_r,v_r)$ of $C_r$ and let $P_r$ be the complementary path from $u_r$ to $v_r$, formed by the remaining $\ell_r-1$ edges. We assign the connection phase $+\alpha$ to the edges of $P_r$ in the direction from $u_r$ to $v_r$. The direct edge $e_r$ is also assigned the phase $+\alpha$ in the direction from $u_r$ to $v_r$, and therefore contributes $-\alpha$ when the oriented cycle is closed by traversing it from $v_r$ to $u_r$.

The accumulated phase around the resulting oriented cycle is
\begin{align}
\mathrm{Hol}_\alpha(C_r)
&=
(\ell_r-1)\alpha +(-\alpha)
\nonumber\\
&=
(\ell_r-2)\alpha
\pmod{2\pi}.
\end{align}
For the chosen cycle orientation, the holonomy of the associated connection therefore equals the path mismatch generated by the symmetric Sakaguchi--Kuramoto lag,
\begin{equation}
\mathrm{Hol}_\alpha(C_r)
=
h_\alpha(C_r).
\end{equation}
Reversing the orientation of the cycle changes the sign of both quantities but does not change the corresponding compatibility or spectral information.

The selected edge is not physically distinguished. Choosing another edge of the same cycle produces the same holonomy, because every direct path contains one edge and every complementary path contains $\ell_r-1$ edges. The resulting connections have the same cycle holonomy and are therefore related by a node-wise gauge transformation.

Because the cycles of a cactus graph do not share edges, this construction can be performed independently on every cycle. Edges that do not belong to any cycle are bridges, and their connection phases may be set to zero by a gauge choice. The collection
\begin{equation}
\left\{
h_\alpha(C_r)
\right\}_{r=1}^{R}
\end{equation}
can therefore be represented by a single global antisymmetric connection, uniquely up to a node-wise gauge transformation.

The symmetric Sakaguchi--Kuramoto lag generates the cycle-level path mismatches
\begin{equation}
h_\alpha(C_r)
=
(\ell_r-2)\alpha,
\end{equation}
and the associated antisymmetric connection represents these quantities as cycle holonomies. Its twisted Laplacian consequently provides a static spectral encoding of the compatibility geometry generated by the symmetric phase lag. As $\alpha$ varies, the cycle holonomies give the exact cycle-resolved mismatches, while the low-lying spectrum provides a graph-dependent description of their combined incompatibility. These geometric markers are not, in general, stability boundaries of the symmetric Sakaguchi--Kuramoto dynamics, but provide spectral markers that can be compared with reorganizations of its phase-locked states.

\subsubsection{Static spectral encoding of cycle compatibility}

Let $L_\alpha=\delta_\alpha^{*}\delta_\alpha$ be the twisted Laplacian of the associated connection constructed in Sec.~\ref{sec:sec2.2}. Ordering its eigenvalues as
\begin{equation}
0\leq\lambda_1(\alpha)\leq\cdots\leq\lambda_N(\alpha),
\end{equation}
the smallest eigenvalue is
\begin{equation}
\lambda_1(\alpha)
=
\min_{\|z\|=1}
\left\langle z,L_\alpha z\right\rangle.
\end{equation}
By the zero-mode condition established above,
\begin{equation}
\lambda_1(\alpha)=0
\end{equation}
if and only if
\begin{equation}
\mathrm{Hol}_\alpha(C_r)
\equiv
0
\pmod{2\pi}
\qquad
\text{for every } C_r\in\mathcal{C}.
\end{equation}
The associated twisted Laplacian therefore provides a static spectral encoding of the cycle-compatibility geometry generated by the symmetric phase lag. A positive value of $\lambda_1(\alpha)$ measures the minimum normalized residual incompatibility after optimization over all node states.

For an isolated cycle $C_r$ of length $\ell_r$, the twisted spectral branches depend only on its holonomy and are given by
\begin{equation}
2-2\cos\left(
\frac{2\pi m+\mathrm{Hol}_\alpha(C_r)}{\ell_r}
\right),
\quad
m=0,\ldots,\ell_r-1.
\end{equation}
Its smallest eigenvalue is therefore
\begin{equation}
\lambda_1^{(r)}
=
\min_{m=0,\ldots,\ell_r-1}
\left\{
2-2\cos\left(
\frac{2\pi m+\mathrm{Hol}_\alpha(C_r)}{\ell_r}
\right)
\right\}.
\end{equation}
It vanishes when
\begin{equation}
\mathrm{Hol}_\alpha(C_r)
\equiv
0
\pmod{2\pi},
\end{equation}
and reaches its maximum when
\begin{equation}
\mathrm{Hol}_\alpha(C_r)
\equiv
\pi
\pmod{2\pi}.
\end{equation}
At this maximally frustrated value, two adjacent winding sectors become degenerate, marking a geometric reorganization of the cycle-compatibility structure. The remote synchronization motif considered next provides concrete three- and five-cycle examples of this spectral transition.

\begin{figure}[!htbp]
\begin{center}
\includegraphics[width=\columnwidth]{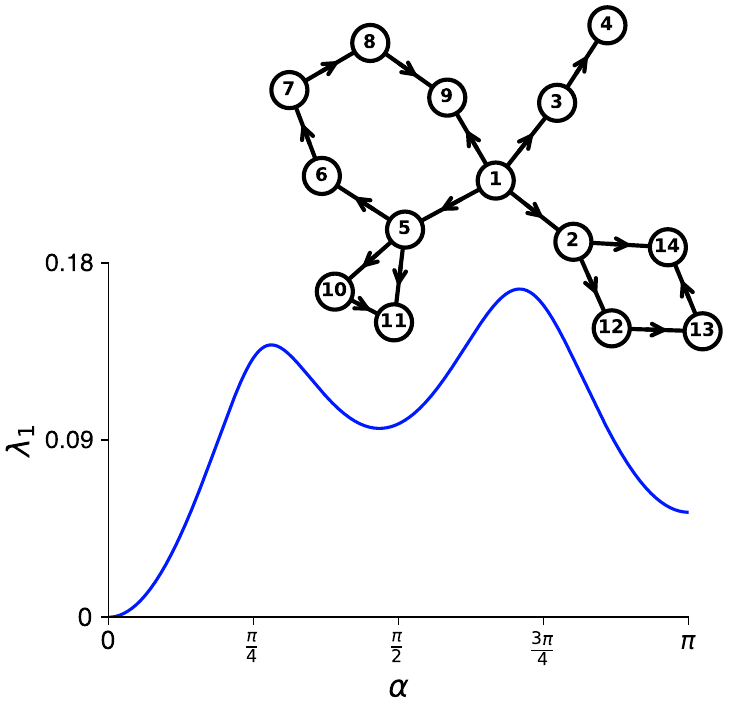}
\end{center}
\caption{\textbf{Global compatibility spectrum of a cactus graph.} Smallest eigenvalue $\lambda_1(\alpha)$ of the associated twisted Laplacian for a cactus containing multiple simple cycles connected by shared vertices and bridges. The curve measures the residual incompatibility of the full holonomy assignment; its extrema reflect the combined effect of the cycle mismatches rather than a sum of isolated-cycle spectra.}
\label{fig:motif2}
\end{figure}

For the full cactus graph, $\lambda_1(\alpha)$ provides a graph-dependent measure of the residual incompatibility generated by the complete set of cycle holonomies. It is not, in general, the sum of the isolated-cycle eigenvalues, because cycles may interact through shared vertices and attached trees. The cycle holonomies remain the exact cycle-resolved quantities, while $\lambda_1(\alpha)$ measures how their combined incompatibility is distributed across the full graph. An example with several cycles and bridge edges is shown in Fig.~\ref{fig:motif2}, where the global curve summarizes the combined compatibility geometry of the associated connection.

\subsubsection{Cycle compatibility in the remote synchronization motif}

We now apply this construction to the remote synchronization motif studied in Ref.~\cite{nicosia2013remote} and used throughout this work. The graph consists of a triangle and a pentagon sharing one vertex. Since the two cycles share no edges, the graph is a cactus and the path--holonomy correspondence applies.

The triangular cycle has length three and carries the mismatch
\begin{equation}
h_\alpha(C_3)
=
(3-2)\alpha
=
\alpha,
\end{equation}
whereas the pentagonal cycle has length five and carries
\begin{equation}
h_\alpha(C_5)
=
(5-2)\alpha
=
3\alpha.
\label{eq:pentagon_mismatch}
\end{equation}
For the chosen cycle orientations, these mismatches are represented by the holonomies of the associated antisymmetric connection,
\begin{equation}
\mathrm{Hol}_\alpha(C_3)
=
h_\alpha(C_3),
\qquad
\mathrm{Hol}_\alpha(C_5)
=
h_\alpha(C_5).
\end{equation}

\begin{figure*}[t]
\begin{center}
\begin{minipage}{0.48\textwidth}
\begin{center}
\includegraphics[width=\linewidth]{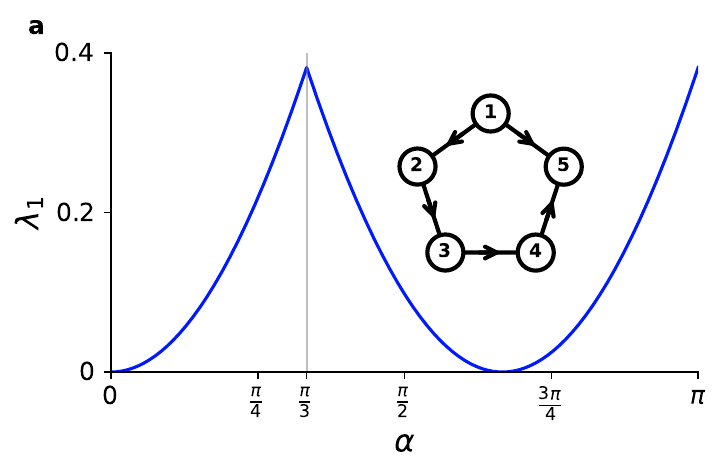}
\end{center}
\end{minipage}
\hfill
\begin{minipage}{0.48\textwidth}
\begin{center}
\includegraphics[width=\linewidth]{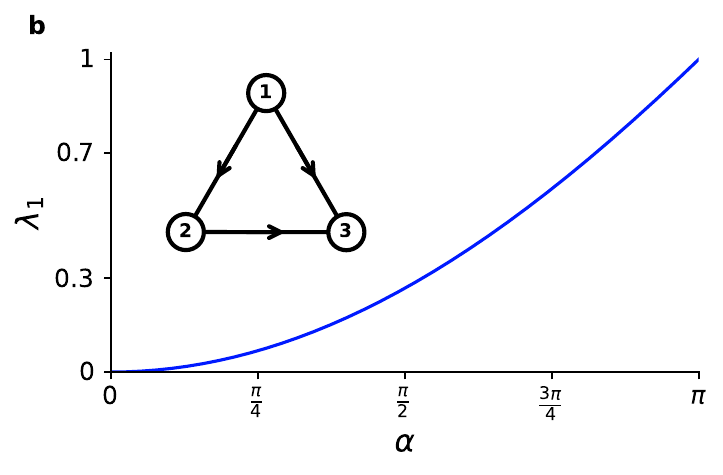}
\end{center}
\end{minipage}
\end{center}
\caption{\textbf{Isolated-cycle compatibility spectra.} \textbf{a} Smallest twisted-Laplacian eigenvalue of $C_5$ for the associated connection with holonomy $\mathrm{Hol}_\alpha(C_5)=3\alpha$. The gray line marks $\alpha=\pi/3$, where the pentagonal holonomy equals $\pi$ and the two lowest winding sectors become degenerate. \textbf{b} Corresponding spectrum of $C_3$, for which $\mathrm{Hol}_\alpha(C_3)=\alpha$; at the same value of $\alpha$, the triangular holonomy remains below its half-flux value $\pi$.}
\label{fig:isolated_cycles}
\end{figure*}

The first positive values of $\alpha$ at which each cycle reaches maximally frustrated holonomy are
\begin{equation}
\alpha_{C_3}
=
\pi,
\qquad
\alpha_{C_5}
=
\frac{\pi}{3}.
\end{equation}
At $\alpha=\pi/3$, the two cycle mismatches are
\begin{equation}
h_\alpha(C_3)
=
\frac{\pi}{3},
\qquad
h_\alpha(C_5)
=
\pi.
\end{equation}
Thus, the pentagonal cycle reaches maximal path incompatibility at $\alpha=\pi/3$, while the triangular cycle remains below its first geometric reorganization point. The pentagon is therefore the first cycle in the motif to reach a degeneracy of its two lowest isolated-cycle spectral branches as $\alpha$ increases from zero. This isolated-cycle contrast is shown in Fig.~\ref{fig:isolated_cycles}; the full motif spectrum in Fig.~\ref{fig:motif} gives the corresponding graph-level encoding.

\subsubsection{Symmetry reduction and the geometric two-group configuration}

Remote synchronization consists of phase-locked clusters containing nodes that are not necessarily adjacent. In the present motif, the admissible cluster subspaces are determined by graph symmetries~\cite{nicosia2013remote,pecora2014cluster,sorrentino2016complete}. The reflection symmetries of the triangle-pentagon motif define the classes
\begin{equation*}
\{1\}, \qquad \{2,3\}, \qquad \{4,7\}, \qquad \{5,6\}.
\end{equation*}
These classes determine the symmetry-related cluster coordinates, including the nonadjacent pair $\{4,7\}$.

Because the motif is not regular, complete synchronization is not a phase-locked solution for a generic nonzero phase lag. If all phases were equal, Eq.~\eqref{eq:SK} would require
\begin{equation*}
\Omega
=
\omega-\kappa k_i\sin\alpha
\qquad
\text{for every node }i,
\end{equation*}
which cannot hold when the degrees $k_i$ differ and ${\sin\alpha\neq0}$. The relevant locked branch is therefore a nonuniform phase configuration within the symmetry-invariant subspace.

A symmetry-respecting phase-locked state can be written as
\begin{align*}
\phi_1(t) &= \Omega t,\\
\phi_2(t)=\phi_3(t) &= \Omega t+x,\\
\phi_4(t)=\phi_7(t) &= \Omega t+y,\\
\phi_5(t)=\phi_6(t) &= \Omega t+z.
\end{align*}
On the branch connected continuously to complete synchronization at $\alpha=0$, the locking equations select
\begin{equation*}
z=y+x.
\end{equation*}
After eliminating the common locked frequency, the remaining equations reduce to
\begin{align}
0={}&
\sin(y+\alpha)-2\sin x\cos\alpha-\sin\alpha,
\label{eq:F1}\\
0={}&
2\sin(x-\alpha)+2\sin(y-\alpha)
+\sin(x+\alpha)+\sin\alpha.
\label{eq:F2}
\end{align}

The locked branch reaches the two-group configuration
\begin{equation*}
x=0,
\qquad
y=z=\alpha.
\end{equation*}
Under this substitution, Eq.~\eqref{eq:F2} is satisfied identically, whereas Eq.~\eqref{eq:F1} becomes
\begin{equation*}
\sin\alpha\left(2\cos\alpha-1\right)=0.
\end{equation*}
The nontrivial solution in the interval $0<\alpha<\pi/2$ is
\begin{equation*}
\alpha=\frac{\pi}{3}.
\end{equation*}
Equivalently, $3\alpha=\pi$. Using Eq.~\eqref{eq:pentagon_mismatch}, this condition becomes
\begin{equation}
h_\alpha(C_5)=\pi.
\label{eq:dynamic_mismatch_equality}
\end{equation}
For the associated antisymmetric connection, this is equivalent to
\begin{equation*}
\mathrm{Hol}_\alpha(C_5)
\equiv
\pi
\pmod{2\pi}.
\end{equation*}

\begin{figure}[t]
\begin{center}
\includegraphics[width=\columnwidth]{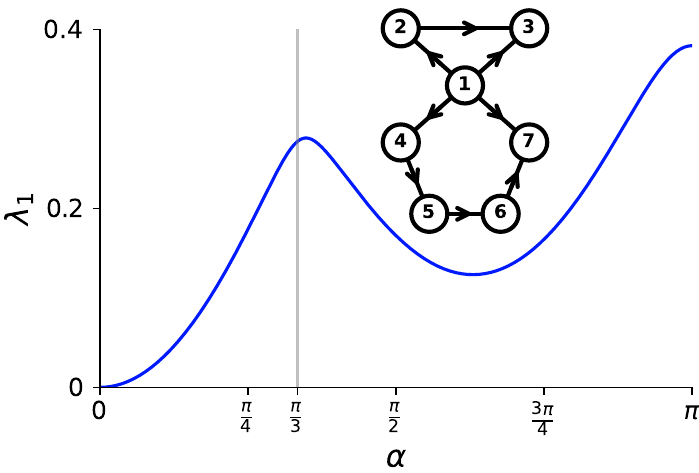}
\end{center}
\caption{\textbf{Compatibility spectrum of the remote synchronization motif.} Smallest eigenvalue $\lambda_1(\alpha)$ of the associated twisted Laplacian for the triangle--pentagon cactus motif. The vertical line marks $\alpha=\pi/3$, where the pentagonal mismatch $h_\alpha(C_5)=3\alpha$ reaches the half-flux value $\pi$ on the isolated pentagonal cycle, while the triangular mismatch $h_\alpha(C_3)=\alpha$ remains equal to $\pi/3$.}
\label{fig:motif}
\end{figure}

Thus, the exact two-group configuration occurs at the same parameter value at which the pentagonal holonomy reaches its maximally frustrated value. The nonlinear locking equations and the isolated pentagonal compatibility spectrum therefore identify the same cycle-resolved geometric point from different descriptions, matching the spectral marker shown in Fig.~\ref{fig:motif}. The holonomy encodes the cycle-level path mismatch, while the locking equations determine how this geometry is expressed in the node phases.

At $\alpha=\pi/3$, the four symmetry classes reduce to two phase groups,
\begin{align*}
\phi_1 &= \phi_2=\phi_3,\\
\phi_4 &= \phi_5=\phi_6=\phi_7
=
\phi_1+\frac{\pi}{3}.
\end{align*}
The physical phase differences still satisfy cycle closure. Hence, $h_\alpha(C_5)=3\alpha$, represented by $\mathrm{Hol}_\alpha(C_5)$ in the associated connection, is not the sum of the physical phase differences. It is the mismatch between the coupling argument along the direct edge and the accumulated coupling arguments along the complementary path.

\subsubsection{Transverse stability of the remote-synchronized branch}

The cycle-compatibility geometry does not determine the stability of the locked branch, which is governed by the Jacobian of the symmetric Sakaguchi--Kuramoto dynamics. Linearization separates the transverse perturbations according to the graph symmetries. The symmetry-adapted Jacobian blocks, together with the branch offsets and transverse eigenvalue with largest real part, are reported in the Supplementary Material. Following the branch obtained from Eqs.~\eqref{eq:F1} and \eqref{eq:F2}, the transverse eigenvalue with largest real part crosses zero in the pentagonal symmetry sector at
\begin{equation}
\alpha_{c}
\simeq
1.0164,
\end{equation}
while the triangular transverse mode remains stable.

This threshold is close to, but distinct from, the geometric maximal-mismatch point,
\begin{equation}
\alpha_{\pi}
=
\frac{\pi}{3}
\simeq
1.0472.
\end{equation}
The maximally frustrated pentagonal holonomy is therefore not a stability criterion. It identifies the point at which the pentagonal path mismatch reaches $\pi$ and the locked branch passes through the exact two-group configuration. Since $\alpha_{c}<\alpha_{\pi}$, this configuration lies on the transversely unstable continuation of the locked branch.

The twisted Laplacian thus encodes the geometric reorganization of the cycle-compatibility structure, whereas the Jacobian determines the stability of the nonlinear phase-locked state.

\section{Discussion}

We have developed a spectral description of cycle compatibility based on the twisted Laplacian. Although phase transport is defined locally on edges, global compatibility is determined by how the prescribed phase shifts accumulate around cycles. This accumulated mismatch cannot be identified from any single edge and acts as an effective higher-order constraint, even though the coupling remains pairwise.

The twisted Laplacian provides a linear representation of this compatibility problem. Its kernel is nontrivial if and only if all cycle holonomies vanish, while its smallest eigenvalue measures the residual incompatibility of the best normalized complex state when the connection is frustrated. For an isolated cycle, the spectrum depends only on the cycle holonomy and reorganizes when the holonomy reaches $\pi$, where the two lowest winding sectors become degenerate.

The gauge-coupled model and the symmetric Sakaguchi--Kuramoto model have different local coupling laws. Their connection arises through path compatibility rather than through a direct identification of the interaction terms. In the symmetric model, the same phase lag appears in both coupling directions, but the coupling arguments accumulated along competing paths may be incompatible. For cactus graphs, these path mismatches can be represented as the holonomies of an associated antisymmetric connection, whose twisted Laplacian provides a static spectral description of the resulting compatibility geometry.

For the remote synchronization motif studied here, graph symmetry determines the admissible synchronized clusters, while the cycle mismatches describe the geometry imposed by the symmetric phase lag. The pentagonal mismatch reaches $\pi$ at $\alpha=\pi/3$, where the isolated pentagonal compatibility spectrum becomes degenerate and the nonlinear locked branch passes through an exact two-group configuration. This geometric point is distinct from the transverse stability threshold, which occurs at $\alpha_{c}\simeq1.0164$ and is determined by the Jacobian of the symmetric dynamics. The cycle holonomies characterize the individual path mismatches, while the global twisted spectrum measures their combined residual incompatibility, the locking equations determine the node phases, and the Jacobian determines their stability. Consequently, no monotone relation between an individual cycle holonomy and the global eigenvalue $\lambda_1(\alpha)$ should be expected in general: $\lambda_1(\alpha)$ is a variational quantity determined by the full graph, and its dependence on $\alpha$ may reflect branch switching, bridges, shared vertices, and the simultaneous presence of several cycle constraints.

More broadly, these results show that higher-order compatibility constraints can arise on ordinary graphs through cycle structure. The relevant information is carried by accumulated path mismatches rather than by individual pairwise interactions. Extensions to heterogeneous phase lags, weighted graphs, and graphs with overlapping cycles may clarify how several coupled cycle constraints combine and how their spectral geometry relates to cluster formation and loss of synchrony.

\section*{Data availability}
No data were used for the research described in this article.

\section*{Code availability}
The code for replicating the results and figures is available at \href{https://github.com/llui2/twisted-laplacian}{github.com/llui2/twisted-laplacian}.

\section*{Acknowledgements}
L.T.-H.\ acknowledges financial support from Diputaci\'o de Tarragona and Universitat Rovira i Virgili, Spain (2023PMF-PIPF-21). This work has been supported by Spanish Ministerio de Ciencia, Innovaci\'on y Universidades PID2024-158120NB-C21, by project PID2022-142600NB-I00 funded by MICIU/AEI/10.13039/501100011033 and FEDER, UE, and Universitat Rovira i Virgili 2025INTER-03 (ComSCIAM) and 2023PFR-URV-00633. A.A.\ also acknowledges the ICREA Academia program from Generalitat de Catalunya.

\section*{Author contributions}
L.T.-H. and A.A. were responsible for conceptualization, methodology, formal analysis, and writing the original draft. L.T.-H., J.D., S.G. and A.A. contributed to reviewing and editing the manuscript.

\section*{Competing interests}
The authors declare no competing interests.

\bibliography{refs}

\renewcommand\theequation{{S\arabic{equation}}}
\renewcommand\thetable{{Supplementary S\Roman{table}}}
\renewcommand{\figurename}{Supplementary Figure}
\renewcommand\thefigure{{S\arabic{figure}}}
\renewcommand\thesection{{Section S\arabic{section}}}

\setcounter{section}{0}
\setcounter{table}{0}
\setcounter{figure}{0}
\setcounter{equation}{0}

\appendix
\section*{Supplementary Material}

This Supplementary Material presents the transverse stability calculation for the remote synchronization motif discussed in the main text. It makes explicit that the geometric maximal-mismatch point and the dynamical stability threshold are close but distinct: the former is determined by the cycle holonomy, whereas the latter is reached when a Jacobian eigenvalue crosses zero along the locked branch.

\subsection*{Transverse stability of the remote-synchronized branch}
\label{sec:transverse_stability}

\begin{figure}[ht!]
\begin{center}
\begin{minipage}{0.48\textwidth}
\begin{center}
\includegraphics[width=\linewidth]{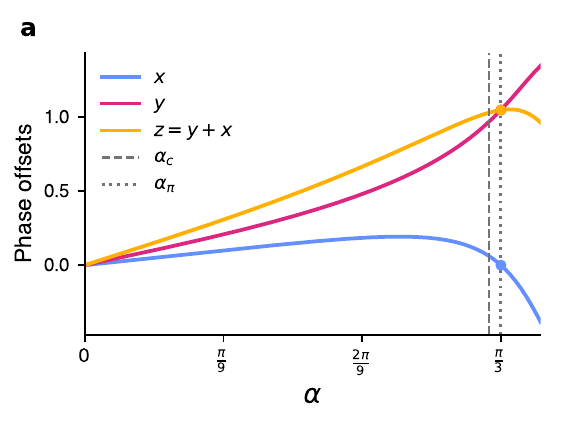}
\end{center}
\end{minipage}
\hfill
\begin{minipage}{0.48\textwidth}
\begin{center}
\includegraphics[width=\linewidth]{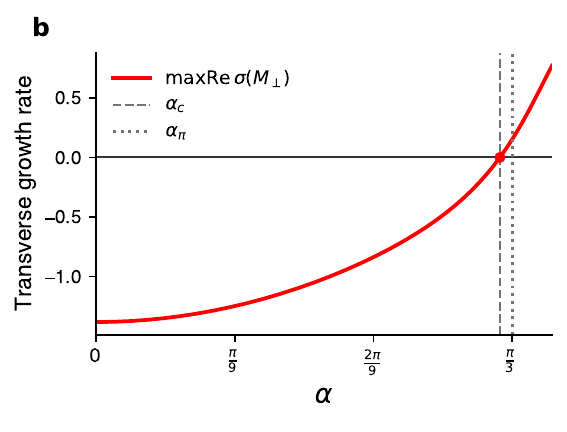}
\end{center}
\end{minipage}
\end{center}
\caption{\textbf{Transverse stability of the remote-synchronized branch.} \textbf{a} Symmetry-reduced phase offsets $x$, $y$, and $z=y+x$ along the branch connected to complete synchronization at $\alpha=0$. The dashed line marks the transverse stability threshold $\alpha_{c}$, and the dotted line marks the first positive geometric maximal-mismatch point $\alpha_{\pi}=\pi/3$. \textbf{b} Largest real part of the transverse eigenvalues, $\max\operatorname{Re}\sigma(M_{\perp})$, in the pentagonal symmetry sector. The zero crossing occurs at $\alpha_{c}\simeq1.0164$, before the maximally frustrated pentagonal holonomy is reached at $\alpha_{\pi}=\pi/3$.}
\label{fig:sm_stability}
\end{figure}

The symmetric Sakaguchi--Kuramoto dynamics is
\begin{equation*}
\dot{\phi}_i
=
\omega
+
\kappa
\sum_{j=1}^{N}A_{ij}
\sin(\phi_j-\phi_i-\alpha),
\qquad
\kappa>0.
\end{equation*}
Since $\kappa>0$ only rescales time and all linearized eigenvalues by a positive factor, we set $\kappa=1$ without loss of generality in the stability calculation below.
The graph symmetries define the four phase classes
\begin{equation*}
\{1\},
\qquad
\{2,3\},
\qquad
\{4,7\},
\qquad
\{5,6\}.
\end{equation*}
A phase-locked state in this invariant subspace can therefore be written as
\begin{align*}
\phi_1(t)&=\Omega t,
&
\phi_2(t)=\phi_3(t)&=\Omega t+x,
\\
\phi_4(t)=\phi_7(t)&=\Omega t+y,
&
\phi_5(t)=\phi_6(t)&=\Omega t+z.
\end{align*}
Equating the locked frequencies of the classes $\{2,3\}$ and $\{5,6\}$ gives
\begin{equation*}
\sin(y-z-\alpha)
=
-\sin(x+\alpha).
\end{equation*}
The branch connected continuously to complete synchronization at $\alpha=0$ selects
\begin{equation*}
z=y+x
\qquad
(\mathrm{mod}\ 2\pi).
\end{equation*}
We use continuous phase representatives for which $z=y+x$.
After eliminating the common locked frequency, the two independent equations are
\begin{align*}
0={}&
\sin(y+\alpha)-2\sin x\cos\alpha-\sin\alpha,
\\
0={}&
2\sin(x-\alpha)+2\sin(y-\alpha)
+\sin(x+\alpha)+\sin\alpha.
\end{align*}
These equations determine the nonlinear locked branch shown in Fig.~\ref{fig:sm_stability}. Along this branch, the first positive geometric maximal-mismatch point of the pentagonal cycle occurs at $\alpha_{\pi}=\pi/3$, where $h_\alpha(C_5)=\pi$. At this value the branch reaches the exact two-group configuration
\begin{equation*}
x=0,
\qquad
y=z=\frac{\pi}{3}.
\end{equation*}

Stability is determined by the eigenvalues of this Jacobian evaluated on the locked branch, not by the holonomy alone. The linearized transverse perturbations separate according to the same graph symmetries. The triangular transverse mode is parameterized by
\begin{equation*}
(\delta\phi_1,\ldots,\delta\phi_7)
=
(0,\eta,-\eta,0,0,0,0),
\end{equation*}
and the transverse modes supported on the pentagonal pairs $\{4,7\}$ and $\{5,6\}$ are parameterized by
\begin{equation*}
(\delta\phi_1,\ldots,\delta\phi_7)
=
(0,0,0,\xi_1,\xi_2,-\xi_2,-\xi_1).
\end{equation*}
The $\eta$ perturbation satisfies
\begin{equation*}
\dot{\eta}
=
\lambda\eta,
\end{equation*}
where
\begin{equation*}
\lambda
=
-\left[
\cos(x+\alpha)+2\cos\alpha
\right].
\end{equation*}
The $\xi$ amplitudes satisfy
\begin{equation*}
\begin{pmatrix}
\dot{\xi}_1\\
\dot{\xi}_2
\end{pmatrix}
=
M_{\perp}
\begin{pmatrix}
\xi_1\\
\xi_2
\end{pmatrix},
\end{equation*}
with
\begin{equation*}
M_{\perp}
=
\begin{pmatrix}
-\cos(y+\alpha)-\cos(x-\alpha) & \cos(x-\alpha)\\
\cos(x+\alpha) & -\cos(x+\alpha)-2\cos\alpha
\end{pmatrix}.
\end{equation*}
We denote the spectrum of this transverse block by $\sigma(M_{\perp})$; thus $\max\operatorname{Re}\sigma(M_{\perp})$ is the largest real part among its eigenvalues.

The first transverse stability threshold is obtained by following the branch connected continuously to complete synchronization at $\alpha=0$ and finding the first positive solution of $\det M_{\perp}=0$. Equivalently, it is obtained by solving
\begin{align*}
0={}&
\sin(y+\alpha)-2\sin x\cos\alpha-\sin\alpha,
\\
0={}&
2\sin(x-\alpha)+2\sin(y-\alpha)
+\sin(x+\alpha)+\sin\alpha,
\\
0={}&
\det M_{\perp}.
\end{align*}
The first solution encountered as $\alpha$ increases from zero is
\begin{equation*}
\alpha_{c}
\simeq
1.0164,
\end{equation*}
with branch values
\begin{equation*}
x_c\simeq0.0631,
\qquad
y_c\simeq0.9657,
\qquad
z_c\simeq1.0288 .
\end{equation*}
At this point, $\lambda \simeq -1.5247$, and the transverse spectrum is $\sigma(M_{\perp})\simeq\{0,-1.7038\}$. The $\eta$ eigenvalue and the nonzero $\xi$-sector eigenvalue remain negative at this point, so the first loss of transverse stability occurs in the pentagonal symmetry sector. The remaining nonzero symmetry-preserving eigenvalues of the full Jacobian are also negative, approximately $-3.0792$, $-0.9704$, and $-0.2285$, together with the neutral eigenvalue associated with uniform phase shifts. Hence this transverse crossing is also the first instability of the full linearized dynamics. The branch values and the transverse eigenvalue crossing, $\max\operatorname{Re}\sigma(M_{\perp})$, are shown in Fig.~\ref{fig:sm_stability}.

The first positive geometric maximal-mismatch point is
\begin{equation*}
\alpha_{\pi}
=
\frac{\pi}{3}
\simeq
1.0472 .
\end{equation*}
Thus,
\begin{equation*}
\alpha_{c}
<
\alpha_{\pi}.
\end{equation*}
The maximally frustrated pentagonal holonomy therefore does not by itself mark the stability loss. It identifies the geometric point at which the pentagonal path mismatch first reaches $\pi$ and the locked branch passes through the exact two-group configuration. The transverse stability boundary occurs when the transverse eigenvalue crosses zero, slightly earlier than this geometric point, so this exact two-group configuration lies on the transversely unstable continuation of the locked branch.

\newpage

\end{document}